\documentclass{egpubl}
\usepackage{EGUKCGVC2024}

\usepackage[pass]{geometry}

\WsPaper           

\CGFccby

\usepackage[T1]{fontenc}
\usepackage{dfadobe}  

\usepackage{cite}  
\BibtexOrBiblatex
\electronicVersion
\PrintedOrElectronic
\ifpdf \usepackage[pdftex]{graphicx} \pdfcompresslevel=9
\else \usepackage[dvips]{graphicx} \fi

\usepackage{egweblnk} 

\title[Creating Data Art: Authentic Learning and Visualisation Exhibition]%
      {Creating Data Art: Authentic Learning and Visualisation Exhibition}
\author[J.C.Roberts]{
\parbox{\textwidth}{\centering
Jonathan C. Roberts\thanks{email: j.c.roberts@bangor.ac.uk}\orcid{0000-0001-7718-3181}
}\\
{\parbox{\textwidth}{\centering Bangor University, UK}
}
}

\renewcommand*{\backref}[1]{
}

\begin{document}


\maketitle
\begin{abstract}
We present an authentic learning task designed for computing students, centred on the creation of data-art visualisations from chosen datasets for a public exhibition. This exhibition was showcased in the cinema foyer for two weeks in June, providing a real-world platform for students to display their work. Over the course of two years, we implemented this active learning task with two different cohorts of students. In this paper, we share our experiences and insights from these activities, highlighting the impact on student engagement and learning outcomes. We also provide a detailed description of the seven individual tasks that learners must perform: 
topic and data selection and analysis, 
research and art inspiration, 
design conceptualisation, 
proposed solution,
visualisation creation, 
exhibition curation, and
reflection.
By integrating these tasks, students not only develop technical skills but also gain practical experience in presenting their work to a public audience, bridging the gap between academic learning and professional practice.


\end{abstract}  
\section{Introduction}
Data art bridges data science, visualisation and art. It is a form of artistic expression that is underpinned by data, or in some cases, data becomes the art form itself.  Individuals engaged in data art aim to create works that are not only visually appealing but also rich in information and meaning. 
Data art opens new ways for people to experience data, creating a platform that can excite and engage the general public. 
By blending creativity with data analysis, it has the potential to not only captivate audiences but raise awareness about various topics.  One recent example is the Bletchley Park exhibition ``The Art of Data: Making sense of the world''~\cite{artOfDataBletchley2024} that showcases various methods the code-breakers used to visualise data, from physical cork boards with string, to 3D, animated, and interactive visualisations. 

Subsequently, the motivation of producing \textit{data art} and the task of creating a visualisation exhibition provide not only a valuable authentic learning experience but also a way to excite students by incentivising them to produce useful and engaging outputs for a course. Authentic assessments involve realistic problems that one might encounter in the workplace. These open-ended questions enable ``students to practice for the complex ambiguities of adult and professional life''~\cite{wiggins1990case}.
This approach encourages students to apply their technical skills creatively, develop their knowledge and engage with the taught content in a practical way.  By working on projects that culminate in a public exhibition, students are motivated to create high-quality work that is both informative and visually compelling. The strategy helps to bridge the gap between theoretical and classroom learning and practical application of skills to a real world problem. Students need to consider not only the practical challenges of mapping data to visual content, including aspects such as size, quality, and resolution, but also how the work will be perceived by the public and how it reflects on the University, both positively and negatively.  In general authentic tasks not only enhances their learning experience but also prepares them for real-world challenges, in our case by emphasising the importance of clear communication and public presentation of complex data.

In this paper, we present a seven-part process that teachers can use to implement a similar authentic learning task. We share our experiences from utilising this task in a Creative Visualisation module, an optional module for third-year computing students pursuing a BSc in Computer Science. The students select a dataset, design a data-art piece, have their artwork professionally printed, and display it in a public exhibition at the end of the year. This exhibition takes place in a prominent public space in the city, offering students a real-world platform to showcase their creations.

\section{Background \& Related work}
We bring over twenty years of experience in teaching computer graphics, visual computing, and visualisation. Our approach combines the technical aspects of computer graphics with graphics modelling and creative visualisation, encouraging students to apply their knowledge to specific challenges. In the ICE3121 ``Creative Visualisation'' module, we integrate the teaching of traditional graphics modelling techniques, such as Lindenmayer systems and Iterated Functions, with creative design methodologies. We use the Five Design-Sheets~\cite{RobertsFdsTVCG2016} framework to structure the design process and incorporate data visualisation mapping techniques.

\subsection{Authentic learning, creative thought and visualisation}
Creating tasks and suitable assessments for students in higher education can be difficult~\cite{biggs2022teaching}. In particular, teachers of third-year and final year students on a computing degree, need to define tasks that are open-ended, yet constrained.  With visualisation and computer graphics units, students are often tasked to develop a solution for a dataset, or visualise a dataset. Students can use different solutions, from programming with D3.js, Python libraries to using processing.org (e.g., see~\cite{LiuETAL2021}). These design tasks encourage students to consider multiple  potential solutions, and develop and apply their skills. They stretch the student's ability, and allow that everyone can find their own solution. However expressing the task in a simple and creative way, that both stretches the individual and engenders critical and creative thought is challenging. 

Teachers should choose appropriate design questions, to draw out the best result in the students, and help them develop, and not be too difficult such to affect anxiety in student performance~\cite{russell2012impact}. Indeed if the task is too open-ended it can frighten the student into doing nothing~\cite{roberts2017five}. Furthermore the grading rubric is harder to define, and if the task is too open-ended it can actually inhibit creativity~\cite{BodenCreativeMind1991}. What is required is a `constrained cognitive environment'; a task that is open-ended, yet is focused to help spark creative-thought by forcing people to work within specific constraints~\cite{bonnardel1999creativity}. 
Fostering creative thinking and critical analysis is essential in higher education. Donnelly~\cite{Donnelly2004} writes ``Creativity and curriculum design is about having and using your imagination and perhaps helping others to use theirs'', going on to explain that fostering creative thinking and getting students to work creatively will empower  them to not only survive but also flourish in the world, leading to more fulfilling and meaningful lives. This emphasis on critical and creative thought is particularly exemplified by the final three stages of Bloom's taxonomy~\cite{krathwohl2002revision}, which guide students to analyse challenges and draw connections between ideas, evaluate problems and justify their decisions, and ultimately design, assemble, and construct an original piece of work. Subsequently, encouraging active participation, and developing authentic assessments, will help empower students with appropriate knowledge and the necessary skills to employ in the workplace.

Therefore, it is crucial for the teacher to carefully consider and choose an appropriate assessment that fosters active participation and offers an authentic evaluation. Various visualisation tasks can be considered as potential options. For instance, asking students to ``create a bar chart'' is too restrictive for a task for higher-level students, such as on a third-year module, because  it only allows a few different variations between students' submissions. Whereas ``choose some data, and design a data-art piece'' is better. The task is open ended, it is `ill-defined'~\cite{RobertsEtAL2014}, not because it is poorly constructed, but because such tasks allow for multiple possible solutions and draws upon creative thinking skills. 

All visualisation designers, whether for visualisation or data-art, need to apply \textbf{creative thought}. People use their imagination to be creative and construct novel and innovative ideas. Artists alike draw on the experiences around them, often interlocking two or more previous (and often unrelated) skills or ideas~\cite{Koestler1964}. Ideas are stimulated, perhaps by divergent thinking~\cite{guilford1956structure}, lateral thinking~\cite{deBonoLateralThinking}. However, novel ideas do not just appear, instead develop because of hard work~\cite{SteveJohnsonGoodIdeas}. The Design Council's Double Diamond model~\cite{council2007eleven} exemplifies this process by highlighting the importance of divergent thinking in the initial phases and the refinement of ideas in the later stages. Consequently students should be provided with an appropriate structure and sufficient time to research relevant work and develop  original design concepts.

For our authentic assessment, we take it a step further by defining the task as follows: ``Create a data art piece using selected data, which will be showcased in an exhibition at the end of the course''.
The latter `for exhibition' statement is important, because it gives focus to the question. It is not only an open-ended question, but is also focused. The focus on the exhibition constrains students to focus their attention. It also provides clear instruction, is brief to explain, yet open-ended to allow creativity, and focused on how a student can judge the quality of the output. The task is \textit{authentic}~\cite{DarlingHammondSnyder2000_Authentic,gulikers2004five}, because it is a challenge that students may be asked to achieve in their job, after leaving higher education. 

Specifying `for exhibition' provides a well-defined set of constraints. We know who the `user' is, when the task is required, what is required (something suitable for display), who will need to understand it (the public), where it will be displayed (on the exhibition wall), and some requirements over the format and resolution. In other words students can discuss and consider clear answers to the 5 W's questions (who, what, how, why, when). 
In our teaching we follow the PASS framework~\cite{RobertsPASS2022}. We \textbf{personalise} the task, such that everyone has their own dataset. It is an \textbf{authentic} task, as the students create a data-art piece for exhibition (plus the label, written description with inspiration and reflection of the work). We separate the task into separate stages, and give \textbf{synchronous} \textbf{activities} to help develop the student towards the goal (including art inspiration and design-sketches using the Five Design-Sheets method~\cite{RobertsFdsTVCG2016}).

\subsection{Data art}
Data visualisation and its specialisations, such as information visualisation and visual storytelling, permeate our everyday lives. We encounter bar charts on television, charts and plots on social media, and various forms of visualisations in workplace presentations. However, data art is less frequently discussed, despite its potential to merge aesthetics with data-driven insights, and despite obvious influences that art and design has on the creation and design of data visualisations. One reason may be that data-art is difficult
to define. Vi\'egas and Wattenberg~\cite{ViegasWattenberg2007} offer one definition: that ``artistic visualisations are visualisations of data done by artists with the intent of making art''. 
Intent is a crucial consideration, but other factors also play a role, including approach, audience, medium, representation of data, and purpose. Each of these aspects can vary in degrees. Trying to explain these nuances, Kosara~\cite{kosara2007visualization} proposed an interpretation that places data art on a continuum ranging from pragmatic visualisation designs, through informative art displays, to artistic visualisations and sublime representations (a similar strategy to Milgram's Virtual Reality continuum~\cite{milgram1995augmented}).
Other dimensions could be considered, such as the spectrum from recognisable to non-recognisable visualisations, the readability of quantitative data, and whether the results have utility and purpose within a specific data domain. We briefly expand these ideas.

The \textbf{intent} of data-art visualisations is an important difference~\cite{kosara2007visualization,Li2018DataArt}. With traditional visualisation, the primary goal is to communicate information clearly, or allow people to interact and explore the data to gain an understanding of the underpinning data~\cite{YiEtalInsights2008}. Designers wish to promote and imply trust, accuracy, effectiveness in communicating quantitative information. The objective is often analytical, aiming to support decision-making, enhance understanding, and provide insights.
With data-art, artists, focus on the beauty, design, narrative and emotional response of the work~\cite{segel2010narrative}. 
The \textbf{design} approach is another difference. Common tools, such as bar charts, line graphs, and scatter plots, are used to help in he clear communication and understanding of the data. Words such as ``scientific'' or ``professional'' are used to explain the design. Whereas data-art artists, emphasise aesthetic beauty, and the emotional impact of the work. The design approach is often more experimental and interpretive,  blending data with creative expression to highlight particular aspects or tell a compelling story.  
The \textbf{audience} is another separating factor. Traditional visualisations often target professionals, analysts or decision makers who wish to interpret data in an accurate way. Visualisations are tailored to meet specific informational needs, providing actionable insights and supporting evidence-based reasoning. Data art can appeal 
to a broader audience, from the general public to art enthusiasts. Viewers engage with the works on a more subjective level, understanding the narrative and emotional context.
\textbf{Medium} and presentation style often differs. Visualisations are commonly presented in digital formats such as reports, dashboards, and interactive tools. Data art is often presented in digital installations, multimedia experiences, and traditional art forms. From sketches~\cite{lee2013sketchstory,RobertsFdsTVCG2016} as exemplified by `Dear Data'~\cite{lupi2016dear}, which prompts the use of hand-drawn data visualisations and showed how personal data can be transformed into meaningful and aesthetically pleasing artwork.  Other artworks use icons and glyphs~\cite{zhang2020dataquilt}, video~\cite{angeslevaCooper}, or photographs. E.g, Obscurity, by the conceptual artist, activist, and hacker Paolo Cirio (\href{https://www.paolocirio.net}{paolocirio.net})  showed blurred photographs of over 15 million police people arrested in the US.
To physical objects and large (12 foot) wall hung pieces, such as the weather artworks by 
Nathalie Miebach (\href{https://www.nathaliemiebach.com}{nathaliemiebach.com}) using paper and rope.

\textbf{Representation, data} and interpretation differs. Representation in traditional data visualisation focuses on accuracy, consistency, and the objective depiction of data and its patterns and trends. Whereas data art allows for more subjective interpretation. Artists may distort the data, interpret it, or abstract it to convey specific perspectives or narratives. In Julie Freeman's work ``Selfless Society'' her data was gathered directly from rodents,  her so called, `Rodent Activity Transmission (RAT)' where her work shows  real-time locations of a colony of mole rats (\href{http://rat.systems/selfless-society/}{http://rat.systems/}). 
In the work by Simon Patterson, named ``The Great Bear'' (1992) -- shown by Vi\'egas and Watternberg~\cite{ViegasWattenberg2007}) -- the visualisation does not depict specific data, but it appropriates a traditional visualisation design; the London Underground tube map. Patterson assigns individual tube lines to represent distinct groups of individuals, such as scientists, saints, philosophers, comedians, explorers, and footballers. While inspired by a visualisation technique (a network diagram) it does not specifically display data, as there is no inherent mapping between data and nodes. This presents one aspect of data-art: art inspired by visualisation techniques. 
In \"Angeslev\"a and Cooper's ``Last Clock''~\cite{angeslevaCooper} they depict a creative analogue clock, which is made from a video feed. As the clock hands move, they leave a trace of what has been happening from camera footage. Jason Salavon's ``The Top Grossing Film of All Time, 1x1'' \cite{Salavon2000} uses a reductionist approach, where he calculated an average colour of each frame of the 1997 blockbuster Titanic, displaying it as a single hue. The visualisation shows the vibrant blues of daytime scenes early in the movie, transitioning through the white tones of the iceberg scenes, concluding with extended segments of darkness as the ship sinks. 
Federica Fragapane~\cite{FedericaFragapane2012} visualised Carbon Dioxide Emissions of annual Co2 emissions in 39 countries, incorporating data from \href{http://stats.oecd.org}{oecd.org} and \href{http://data.worldbank.org}{data.worldbank.org}.

\begin{figure}[ht]
    \centering
    \includegraphics[width=\columnwidth]{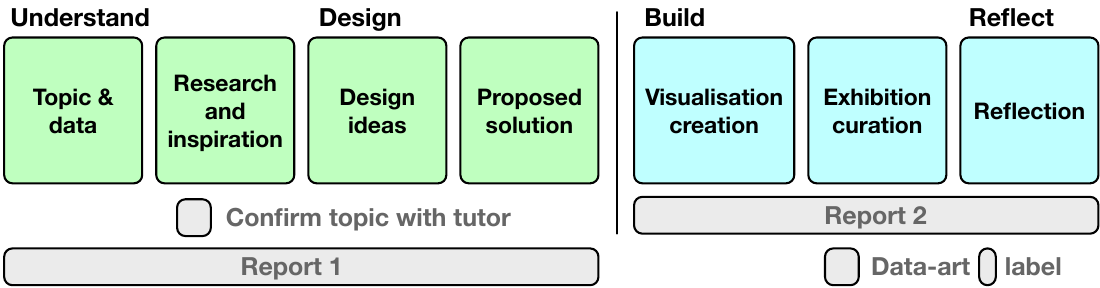}
    \caption {For the exhibition's authentic task, we divided the students' work into two phases: first, to understand the topic and create suitable designs; second, to build the solution, deliver it for the curated exhibition, and reflect on their accomplishments.}
    \label{fig:process}
\end{figure}

\section{Teaching materials and lectures}
We develop the exhibition for the Bangor University module {ICE3121} titled `Creative Visualisation'.  
Creative visualisation brings together creative practice, design, computer graphics and computer algorithms. We explain to students that they will learn to craft digital visual experiences, namely the data-art. Using the creative workflow students aim to learn a suite of processes, that will equip them in a variety of roles in the digital industry. The module features the use of the Five Design-Sheets (FdS) sketched design methodology~\cite{RobertsFdsTVCG2016}. The aims of the module are (1) To present details of creative methods, technologies and visualisation for the purpose of storytelling and creating explanatory visualisation and developing rich experiences. (2) To allow students to develop their programming and creative skills and make consideration of alternative designs, especially using sketching and low-fidelity techniques, in the context of creativity and visualisation. (3) To enable students reflect and critique their work, and how their work in explanatory, creative computing, storytelling and graphical systems, fits with the world. The ICE3121 module is an optional module, open to all computing students in the school. It takes place in the final Semester of their degree. This means that the exhibition is one of their final events, as it occurs after all the final exams.

At the beginning of the course we show the students the public exhibition space (which is front of house of our cinema, theatre and innovation hub). We explain about data-art and enthuse the students about the exhibition, and start to focus their minds onto the requirements of the output. Specifically, students will need to produce high-resolution outputs that exceed the quality of a typical screenshot. We get students to use \href{htpp://processing.org/}{processing.org} and render their vector-based visualisations to PDF. We explain that the works will be printed professionally, with costs being covered through the teaching budget. We believe it is crucial to clarify what students are expected to deliver. Understanding the final output and its requirements helps to focus their efforts and guide their work effectively. Swaffield~\cite{Swaffield2011_HeartOfAuthenticAssessment} refers to this process as `conceptualising the objectives'.

We separate the work, that students are expected to do for the module, into two general stages, with seven overall tasks. Figure~\ref{fig:process} provides a schematic of the  stages and sections of this process. In the first stage, students prepare and make plans. They choose their topic, necessary datasets, explore alternative design ideas, and propose a final solution (that they will create and deliver). Ongoing feedback is given to the students each week, and formal feedback and a grade is given for the first stage. We meet with each student at this stage, and make sure that they know what they are doing, and suggest changes or adaptations to the design ideas. They make changes as appropriate. They present this work in a 1400 word report in the IEEE conference format. 
The second part gets the student to focus on the creation of the data-art work, and deliver the materials for exhibition, plus write a descriptive report of their achievements, techniques used and a reflection of their work. This is submitted as a second report, along with the final artwork, and exhibition label.

\section{The seven tasks}
In this section, we outline each phase and describe the various lectures and activities we use to guide the students.

\textbf{Topic and data.}
First students choose their topic, and start to do some research. We give an introductory lecture explaining the structure of the module, and many data-art examples. We allow the students to choose their own datasets, but provide several examples.  E.g., sports games, Wimbledon winners, football managers, bird population, or city temperatures. In past assessments~\cite{RobertsPASS2022} we have provided a list of topics and datasets for students to choose from. Instead, we encouraged students to select topics that interested them personally, based on their hobbies, political views, or backgrounds. The goal was to increase motivation by allowing them to work on ideas they were already passionate about. For example, one student chose to analyse the frequency of play of different Pokémon, another focused on birds, and another on data of missing people in their country.

Students were given a lecture on data formats, data analysis and data structures. From this lecture they were tasked to research their topic and data. They had to demonstrate a specific underpinning dataset (usually in excel format). On the second week, and before proceeding too far with their projects, students needed to `register' their data topic and dataset. They also had a teacher-consultation. This step is crucial not only to confirm that the data is suitable for the intended analysis but also to ensure that the topic aligns with the project's goals and reflects positively on the university, especially for public exhibition. 
For instance, one student initially wished to focus on the cost of drugs, while another wanted to explore the topic of weapons. Both students were guided to change to different topics that would be more relevant and positively reflect the university's values and reputation. This early consultation helps students to avoid potential pitfalls, ensuring their efforts are focused in a productive direction and that the resulting projects are appropriate for public display. This process underscores the importance of selecting topics that are both academically rigorous and suitable for the intended audience, thereby enhancing the overall impact of the exhibition. 

At the end of the whole first stage, students need to submit their first report. The data and topic is the first part of this report. Consequently students were encouraged to describe their data, and write up the work as they proceeded. They were encouraged to explain the topic and background information on the data. How the data was captured, stored, saved, range of data points, and so on. For instance, they were encouraged to consider who had created it, where was it first published, who has used it, what examples are in use today. Also to explain the potential of the data and consider what information could be displayed, and the story that may be explained. If relevant students were encouraged to include detail of how to process the data (average, sample, categorise, filter, parse) and how it could be used.

\begin{figure}
    \centering
    \includegraphics[width=\columnwidth]{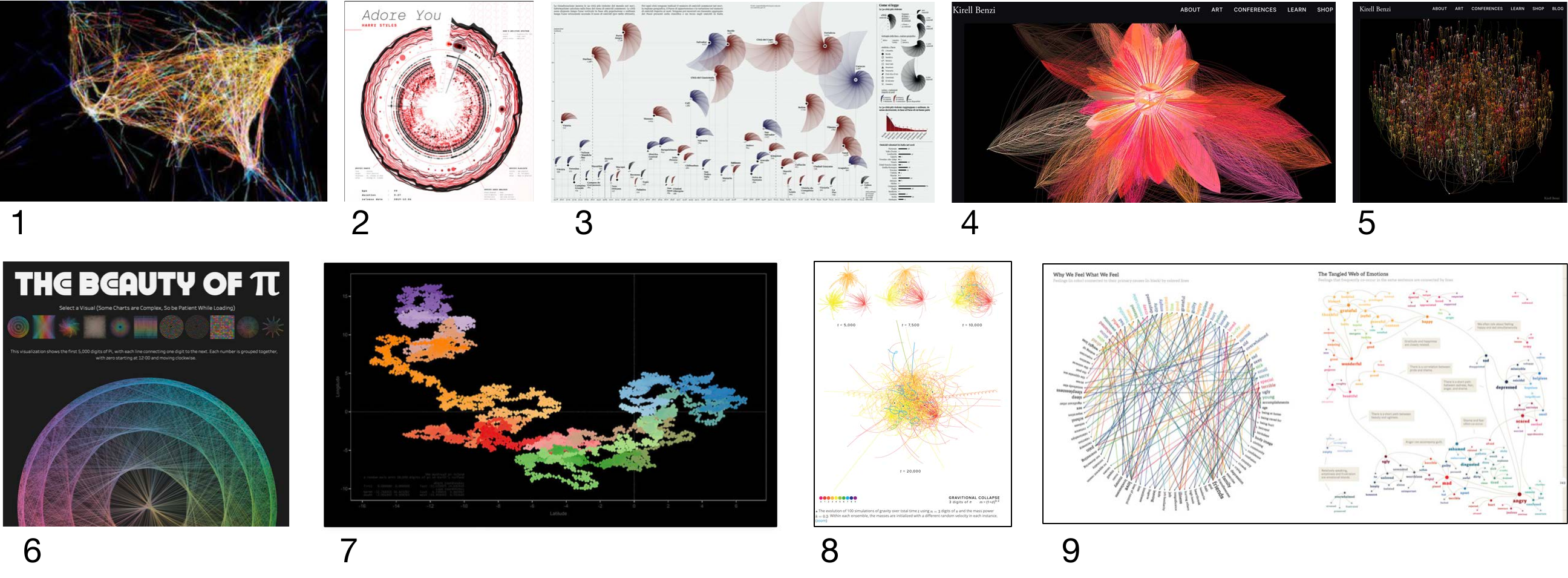}
    \caption{Data art examples. See Appendix 1 for full credits. 1) Aaron Koblin, 2) Nadieh Bremer (2020).
3) Federica Fragapane (2009),
4) Kirell Benzi (2016),
5) Kirel Benzi (2020).
6) Ken Flerlage. 
7) Cristian Vasile (2014) 
8) Martin Krzywinski (2016)
9) Jonathan Harris and Sepandar Kamvar.}
    \label{fig:DataArtExamples}
\end{figure}
\textbf{Research and inspiration.}
We start by giving lectures on (i) general art, (ii) specific data-art examples and (iii) algorithms and mapping techniques to create the artworks. We then give students an in-class activity, which is a scavenge hunt. Where they locate different artists' work, and art styles. For their own project, each student is encouraged to research art movements, including modern, contemporary, impressionist, abstract, cubisim, and so on. The lectures cover examples of artists' work (Figure~\ref{fig:DataArtExamples} shows some of the examples we use). We ask the students perform their own research. To discover the work of a variety of artists, and write up the work in their report. They should locate artists and works that can be cited, especially focusing on artists with publications, blogs, installations, performances and exhibitions. We also lecture about techniques to display the data. We detail data-mapping, layout strategies, and focus on a variety of mapping and graphics algorithms that they could apply to their work. In these lectures we include dense pixel displays, such as pixel bar charts~\cite{KeimPixelBarCharts2002}, iterated function systems (IFS), such as Barnsley fern's~\cite{barnsley1985iterated} and Lindenmayer systems~\cite{lindenmayer1974adding}, point based and growth models including Diffusion Limited Aggregation (DLA)~\cite{witten1983diffusion}, and linear mapping (lerp), non-linear, colour maps and grid layouts.

Students compile their research and inspirational ideas into a dedicated section of their report. They are encouraged to treat this section as a `mood board' of concepts, locating and referencing original works, with a focus on publications, journals, papers, books, exhibitions, and similar sources.

\textbf{Design Ideas}.
To consider the design ideas, we give a lecture on the Five Design-Sheets~\cite{RobertsFdsTVCG2016} and an interactive design activity, on sketching many different alternative ideas. In this group activity they are asked to sketch an idea themselves, twenty seconds, and then share it with the group. They get points if their idea is different to their peers. This exercise encourages them to think beyond common solutions and to explore different ideas from those of their fellow classmates. The Five Design-Sheet method leads students to think through their ideas, and refine them into a final idea. On sheet 1, they sketch many solutions. Sheet 2,3 and 4, allows them to focus on three alternative ideas in more detail. Finally, on sheet 5, students describe their proposed solution and include this write-up in their report. They scan their Five Design-Sheets and add them as appendices, while incorporating elements of their design solutions as figures within the main body of the report.

\textbf{Proposed solution.}
The last part of the first stage is to explain the proposed solution in their report. In other words, this part of the report presents and explains ideas from sheet~5 of the Five Design-Sheet process.  Students need to explain what they will create, what it will look like, and how it will display  data. This represents the final part of their report. They submit the report about one third into their taught content. 
The report is graded, and the teacher provides feedback on their ideas in the next class. This feedback is crucial as it offers students clear validation of their work and guidance, especially if they are straying from the exhibition goal.

\textbf{Visualisation creation.}
We get the students to use \href{http://processing.org}{processing.org}. We give lectures and get the students to run through a suite of exercises. We have students use Processing.org for several compelling reasons. Processing is an excellent platform for teaching and creating visual art and design projects because it is specifically tailored for visual programming. It is a straightforward yet flexible environment, ideal for beginners, yet powerful enough for advanced users. One key advantage is that it can easily render high-resolution vector PDF files. This is crucial for the data art challenge. It ensures that students' work can be scaled to various sizes without loss of quality. Processing.org is widely used in both educational and professional settings, and we have many years of experience of using it, and it provides students with valuable skills that are relevant to work.

\textbf{Exhibition curation.}
The final artwork is submitted by the students in a PDF file, which is then printed professionally. They also create text for the exhibition label. Each label has a \textbf{title}  -- a short descriptive title of their data art, an author \textbf{credit}, which is their name,  and a \textbf{description} of their artwork in 120 words or less. In their description they must include three paragraphs.  Paragraph 1 describes and introduces the work, explaining what is displayed. Paragraph 2 explains the data, and topic, and and what it represents. Paragraph 3 explains how the data is mapped. The final section of the label description includes acknowledgements of sources, images, and libraries used. Students are also encouraged to include different versions of their work. They can submit a folder containing alternative visualisations, which might incorporate different datasets, initial and final design concepts, or variations such as different colour schemes. 

\textbf{Reflection.}
In their reflection report, students should explain the results and include screenshots of their data art. They need to highlight their key achievements and describe the techniques they used. Additionally, they should provide code snippets to illustrate their key techniques and algorithms. Finally, they include a critique and reflection of their work. Aspects that they have achieved, or parts that they did not manage to complete. They reflect on their process, and how they were inspired from their art research. Finally, future ideas of what they could achieve. 

\section{Scenario and exhibition}
To give students a live demonstration, we developed two scenarios that serve as practical case studies to illustrate the concepts. First on world happiness data, and the second on sounds, which we briefly explain below, followed by the final exhibition.

The World Happiness Report~\cite{helliwell2024world} is a dataset that tries to explain the happiness of the world. The data is achieved through several Gallop polls. It details how people self-assess their life satisfaction. One of the main metrics used is the Cantril Ladder. Participants of the survey are asked to imagine a ladder where the top rung (rated as 10) represents the best possible life, and the bottom rung (rated as 0) represents the worst possible life. They then rate their current life satisfaction on this scale. There are notable differences between countries, with recent data showing European nations at the top. Finland, Denmark, Iceland, Switzerland, and the Netherlands have the highest scores, while countries like Afghanistan, South Sudan, and several central Sub-Saharan African nations have the lowest scores.

The inspiration for our data-art came from Chernoff faces~\cite{chernoff1973use}. Chernoff face glyphs were created to present multivariate datasets. While there is discussion over their efficacy~\cite{morris2000experimental}, they can be used to present and interesting data-art piece. However, the typical Chernoff faces often look like distorted eggs, and lack visual appeal. Our aim was to create a design that was more attractive, cute, and artistic than the typical Chernoff faces. We opted for a circular arrangement with a sketch-rendered appearance. 

To create the face visualisations, we refined the data by focusing on 160 countries from the 2024 World Happiness data. With data-art we wish to focus on the general story, at the expense of depicting every data point. In fact, the original 2024 data covered 165 countries. But there is a challenge with this quantity. First many fields are missing in the 2024 data. Consequently we supplemented the missing data with equivalent values from the previous years.  Nonetheless, we still had missing data from Cuba, Somaliland, South Sudan, and the State of Palestine, and merged Congo (Brazzaville) and Congo (Kinshasa), leaving us with 160 countries. Furthermore, the quantity of 165 does not fit well in a grid. We wished to have a balanced grid of faces. Yet with 160 it would allow us to have several grid forms. Configurations could be 16 rows by 10 columns, 20 rows by 8 columns, or 40 rows by 4 columns. 

\begin{figure}
    \centering
    \includegraphics[width=\columnwidth]{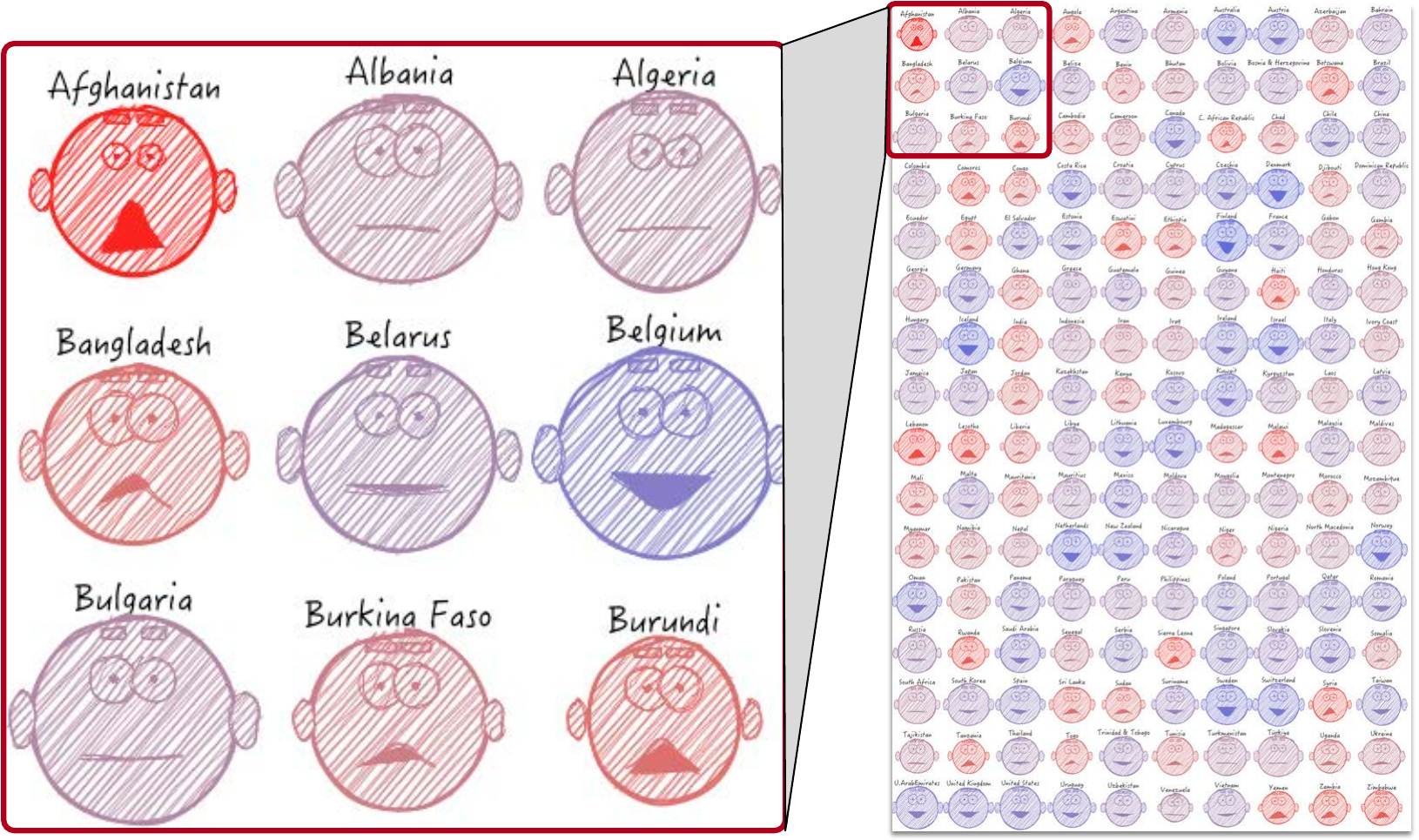}
    \caption{World happiness data, shown in Chernoff face style glyphs, using processing.org. The overall happiness value corresponds to both the depth of the smile and the facial colour. Social support perception is represented by smile width, generosity is indicated by brow length, GDP is reflected in the overall face size, and life expectancy is shown through ear length. Design, data processing and artwork by author.} 
    \label{fig:worldHappinessFaces}
\end{figure}

Using Processing.org, we designed the overall shape and adjusted variables like eyebrow length and ear shape. To achieve the rendered look, we used the Handy processing library~\cite{WoodSketchy2012}. This library gives a sketch-like appearance and allows for PDF output. We experimented with different orderings and found that an alphabetical arrangement was easiest for locating countries and enhanced the visual appeal of the artwork. It also gave variety to the display. For instance, organising the faces by the main Cantril ladder `happiness' score, gives a less interesting data-art piece.
Figure~\ref{fig:worldHappinessFaces} shows the final rendering. The degree of overall happiness corresponds to both the depth of the smile and facial coloration. Social support perception is represented by smile width, generosity is reflected in brow length, GDP correlates with the overall size of the face, and life expectancy is indicated by ear length.

\begin{figure}
    \centering
    \includegraphics[width=\columnwidth]{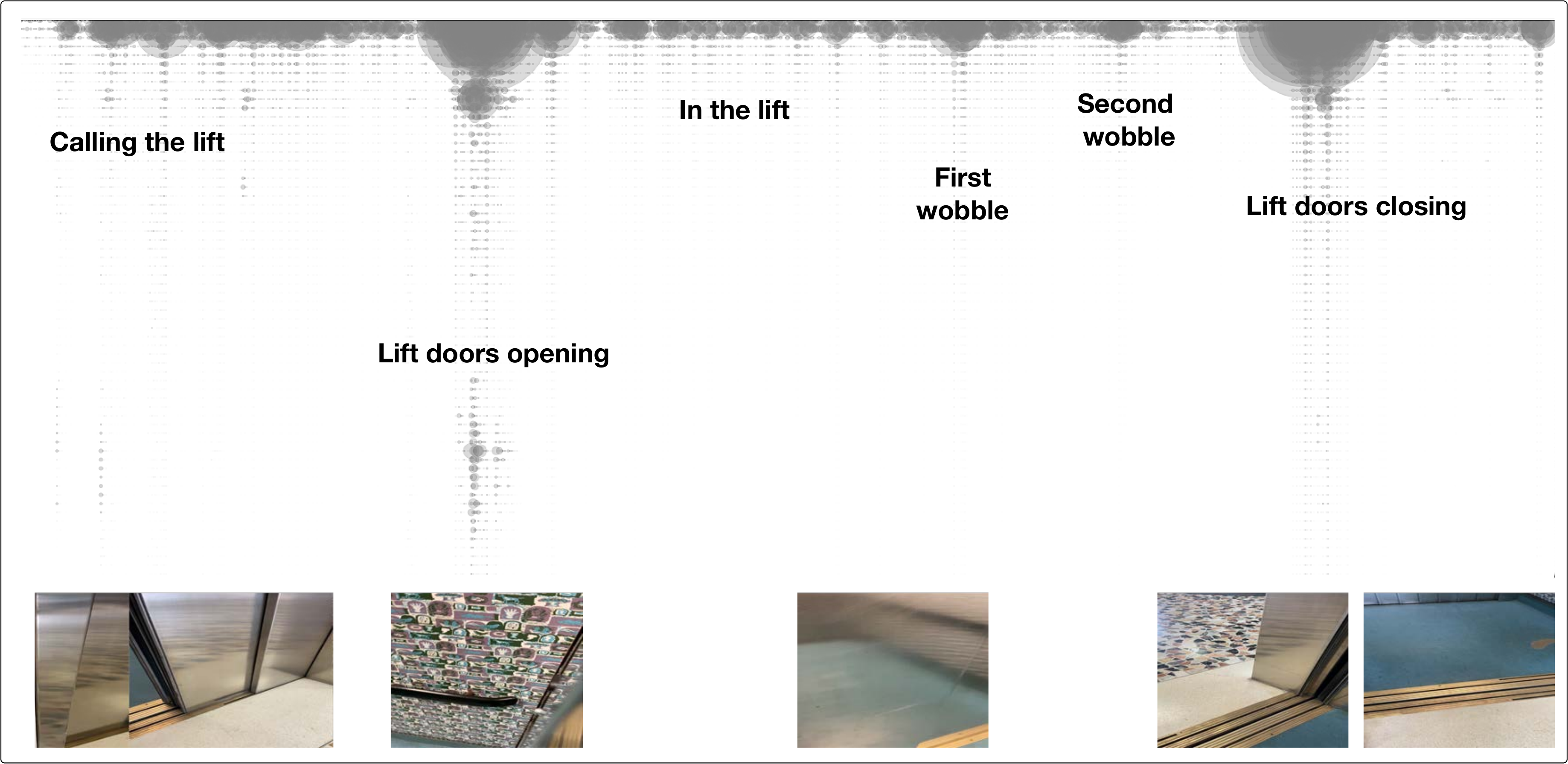}
    \caption{The soundscape data-art work, depicts sounds from the environment. This figure shows a lift being called, travelling in the lift, two wobbles as the lift nears its floor, and the doors opening. Design, data processing and artwork by author.}
    \label{fig:FFTlift}
\end{figure}
The second example is `soundscapes'. We take different recordings of sounds of doors opening, windows opening and closing and so forth. We use the Fourier Transform (FFT) analyzer in processing to calculated the spectrum of the audio stream. Our data art combines the spectrum picture and snapshots from the video stream, see Figure~\ref{fig:FFTlift}. While this is a simple mapping, it provides a useful example of mixing video with sounds and visualisation.

\begin{figure}
    \centering
    \includegraphics[width=\columnwidth]{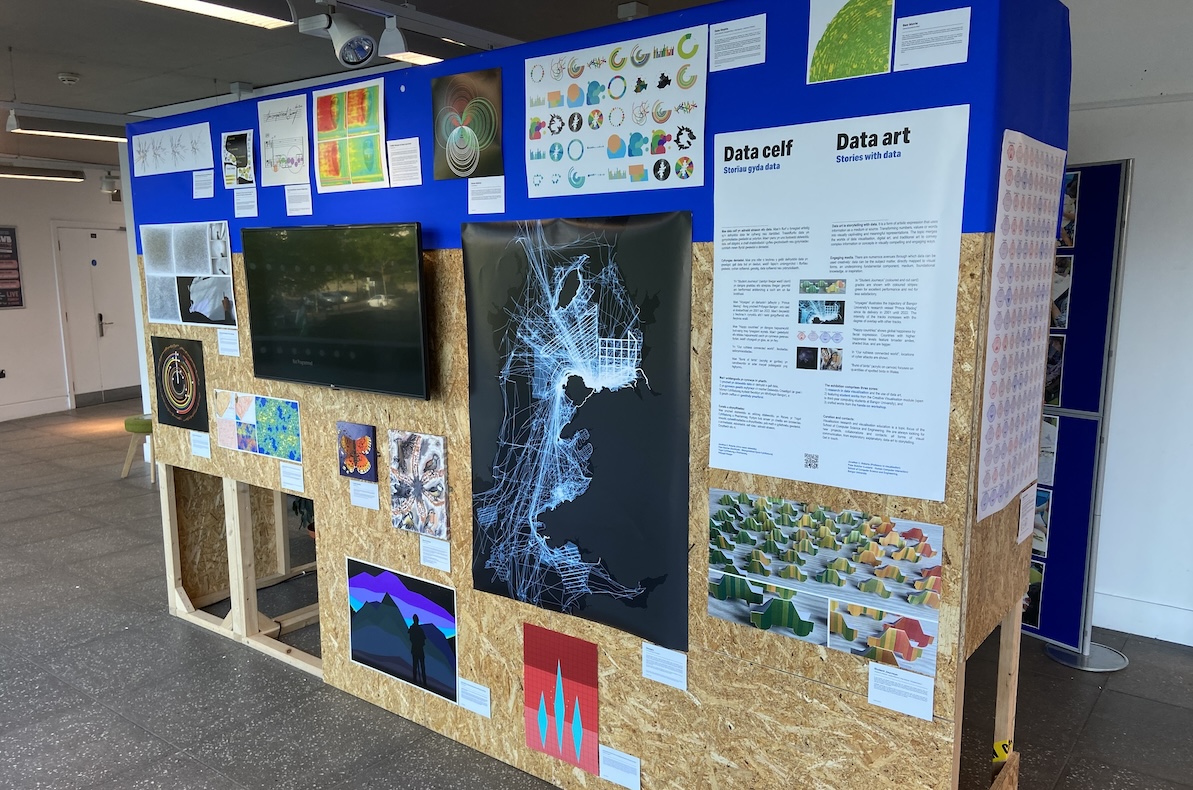}
    \caption{One part of the exhibition, showing some of the exhibits in the foyer of the public innovation, cinema and theatre building in Bangor, UK.}
    \label{fig:Exhibition}
\end{figure}

The 2024 exhibition (Figure~\ref{fig:Exhibition}) showcased 33 unique works created by students and researchers as part of their authentic assessment. It offered the public an insight into the course and student work, data art and visualisation. Each work was presented with a label description, and was shown for two weeks, where visitors could engage with the intersection of art and data science. One of the students said that they were thrilled to see their ideas materialise into tangible artworks. The topics ranged from missing people in Mexico, inter University varsity games between our University and another local University, popular birds in the UK, bird sounds, popularity and use of different Pokemon characters in games, space flights, to the life and death of characters in Tolkien's Lord of the Rings.

\section{Discussion and Conclusion}
The data art exhibition served as a culmination of the authentic learning experience for students. It enabled the students to showcase their creativity and technical skills. They were able to choose and select a dataset, and transform their ideas into meaningful narratives. 
During their studies, students were challenged to conceptualise and design their artworks. Starting the lectures with the vision and goal of creating a data-art exhibition helped to focus the minds of the students on their task. It helped them understand the requirements and expectations of the course. They were able to also consider issues of data accuracy, data analysis, and especially how their work could be viewed by members of the public. The goal of the exhibition provides a useful real-world scenario, and consider some of the complexities of how to create suitable work.

The exhibition demonstrated the variety and breadth of data art. While some students concentrated on traditional visualisation techniques, the majority explored innovative and creative ways to present their data. Indeed, one student developed a visualisation that appeared similar to a scatter plot, and another produced a series of bar charts. However, most students wanted to create something new and realised that the output was different to a typical data-visualisation dashboard. Furthermore, because every student had a different dataset, it meant that they could help each other, and learn from each other. In a few classes, it was clear that the students came into the class and shared with each other what they had achieved that week. It was wonderful to see that they were helping each other by providing feedback. We heard one student suggesting to another that they should remove the separate legend, and try to integrate the key into the main visualisation. They discussed how this could be achieved.  

The delivery of the taught content, with the lectures followed by activities worked well. It enabled the students to try skills directly after the lecture. In Higher Education, a challenge lies in terminology that implies specific modes of operation. We commonly use terms like lectures, classes, and laboratories, each implying a specific way of working. Similar to other institutions, our timetable must specify whether the teaching event, pertains to a lecture or laboratory session. Using terms like `workshops' would align more closely with our educational approach, blending didactic, explanatory, interactive, and reflective learning experiences. The activities also helped the students to interact with each other, and that they able to speak and help each other. The breadth of the lecture material seemed useful. It was clear from the students' reports that they had researched and understood the variety of method, artists, and had explored many different data-artists. However the inclusion of the algorithms lectures, seemed to split the students. Some students, as shown in how they presented their work in their report, demonstrated that they clearly understood that there were different ways to map the data to the display. Whereas about a third of the students missed explaining different algorithmic mappings, instead focusing on the art inspiration and design aspects of the data-art. 

Our chosen methodologies proved effective. Using the Five Design-Sheet method helped students to explore alternative ideas and articulate them clearly. It also provided students with a structured approach that encourages independent research, and they can gain help from other students who have used the same method. 

The feedback and engagement from visitors underscored the impact of the authentic task. The feedback was positive and encouraging. They focused on the variety and diversity of the works, and the professionalism of the individual artworks. Their feedback reinforces the value of the authentic assessment, as the students gained invaluable experience in project management, collaboration, and presentation skills. Essential competencies for their future careers in data science, design, and beyond.

With every assessment and task, that involve students, there are challenges. One challenge is that it does take longer to organise than a traditional module. Setting deadlines in advance, arranging exhibition spaces, and receiving and preparing students' work for printing are essential tasks. Despite clear instructions, students may not always adhere to them as expected. All students submitted something for display. But it took much effort by the tutor, delivering extra help sessions, to get the remaining few struggling students through. One student, meeting with the tutor about another project, expressed pride in being part of the ``show'' yet they had not submitted their work weeks after the printing deadline had elapsed, and only a few days remained before the final exhibition setup. Their work was printed at the last minute and successfully displayed in the exhibition. One student had a label that far exceeded the 120 word requirement, which needed to be edited. And another student submitted a screenshot of their result (ie., a low quality production). However they did submit their processing code. Subsequently the tutor added in the vector  instructions in the code, and rendered the students work into PDF, for high quality printing. While their graded work reflected their mistake, the exhibition display was produced at high quality.

Ultimately, the exhibition not only celebrated the culmination of the students' academic journey but also affirmed the relevance and importance of integrating authentic learning experiences into education. As academics we need to find and use suitable ways that help students understand the challenges that they may face in the workplace. It is not always easy to consider new ways to create tasks that help to drive the students forward, motivate them and give them clear direction and experience that they may find in the workplace. The goal of the data-art exhibition fulfils these goals. It helps to reinforce the idea that learning goes beyond theoretical knowledge to encompass practical application and creative expression. It helps to equip the students with with the confidence and capability to tackle real-world challenges in an innovative and effective way. 
\bibliographystyle{eg-alpha-doi} 

\bibliography{data-art-exhibition}  



\section*{Appendix 1 -- Image credits for Figure~\ref{fig:DataArtExamples}}

1) Aaron Koblin (2009), Flight Patterns, \href{https://vimeo.com/5368967}{https://vimeo.com/5368967}

2) Nadieh Bremer. (2020) Data from the song Adore You by Harry Styles, \href{https://www.visualcinnamon.com/2020/06/sony-music-data-art/}{https://www.visualcinnamon.com/2020/06/sony-music-data-art/}

3) Federica Fragapane, (2009) Artwork for Visual Data, the column on ``La Lettura'',
the cultural supplement of ``Corriere Della Sera''. Depicting carbon dioxide emissions, \href{https://www.behance.net/gallery/31279439/Carbon-Dioxide-Emissions}{https://www.behance.net/gallery/31279439/Carbon-Dioxide-Emissions}

4) Kirell Benzi. ``Secret Knowledge'' (2016) - Wikipedia data. Each node represents a group of pages visited by 
large numbers of people on a specific subject related to world news or events. 
\href{https://www.kirellbenzi.com/art/secret-knowledge}{https://www.kirellbenzi.com/art/secret-knowledge}

5) Kirel Benzi ``Together'' (2020). 
This artwork visualises the hierarchical positions of employees in the Havas group. 
\href{https://www.kirellbenzi.com/art/together}{https://www.kirellbenzi.com/art/together}

6) Ken Flerlage. The beauty of Pi. Inspired by the Pi art of Martin Krzywinski and Cristian Ilies Vasile. For more information see \href{http://mkweb.bcgsc.ca/pi/art/}{http://mkweb.bcgsc.ca/pi/art/}, and \href{https://fineartamerica.com/profiles/cristian-vasile}{https://fineartamerica.com/profiles/cristian-vasile}

7) Cristian Vasile (2014) ``Random Walk of Pi \#1''. \href{https://fineartamerica.com/featured/random-walk-with-pi-1-cristian-vasile.html}{https://fineartamerica.com/featured/random-walk-with-pi-1-cristian-vasile.html}

8) Martin Krzywinski (2016),  ``Gravatational collapse''. The evolution of 100 simulations of gravity over total time. 
\href{https://mk.bcgsc.ca/pi/piday2016/methods.mhtml\#l2home}{https://mk.bcgsc.ca/pi/piday2016/methods.mhtml\#l2home}

9) Jonathan Harris and Sepandar Kamvar, ``Poetic Flows: We Feel Fine'', \href{http://www.wefeelfine.org}{http://www.wefeelfine.org}. Collecting Internet sentences beginning with `I feel' or `I am feeling'. Nearly 10 million feelings and more than two million blogs were collected.

\end{document}